\documentclass[11pt,a4paper]{article} 
\pdfoutput=1
\usepackage{jcappubMOD}
\usepackage{latexsym,amsmath,amssymb}
\usepackage{graphicx}
\usepackage{slashed}
\usepackage{bm}
\usepackage{epsfig}
\usepackage{dcolumn}

\newcommand{\exclude}[1]{}

\newcommand{\beq}{\begin{eqnarray}}
\newcommand{\eeq}{\end{eqnarray}}
\newcommand{\be}{\begin{eqnarray}}
\newcommand{\ee}{\end{eqnarray}}
\newcommand{\bea}{\begin{eqnarray}}
\newcommand{\eea}{\end{eqnarray}}

\newcommand{\rar}{\rightarrow}
\newcommand{\Rar}{\Rightarrow}

\def\dd{\text{d} }

\def\+{\dagger}

\def\<{\langle}
\def\>{\rangle}

\newcommand{\cph}{\varphi}

\newcommand{\cH}{{\cal H}}
\newcommand{\cA}{{\cal A}}

\newcommand{\cL}{{\cal L}}

\newcommand{\Mpc}{\text{Mpc}}
\newcommand{\Gauss}{\text{Gauss}}
\newcommand{\V}[1]{{\boldsymbol{#1}}}

\newcommand{\Vk}{\V{k}}

\newcommand{\Vq}{\V{q}}

\newcommand{\Veps}{\boldsymbol{\varepsilon}}

\newcommand{\mpl}{m_{\rm P}}

\title{Resonant magnetic fields from inflation}
\subheader{
\ifpdf\href{http://dx.doi.org/10.1088/1475-7516/2012/03/009}{JCAP 1203 (2012) 009}\else{JCAP 1203 (2012) 009}\fi
\\ CERN-PH-TH/2011-279
}

\author[a]{Christian~T.~Byrnes,}
\author[b]{Lukas Hollenstein,}
\author[b]{Rajeev Kumar Jain}
\author[c]{and Federico~R.~Urban}

\affiliation[a]{CERN, PH-TH Division, CH--1211, Gen\`eve 23, Switzerland}
\affiliation[b]{D\'epartement de Physique Th\'eorique and Center for Astroparticle Physics, Universit\'e de Gen\`eve, 
24, Quai Ernest Ansermet, CH--1211 Gen\`eve 4, Switzerland}
\affiliation[c]{Department of Physics \& Astronomy, University of British Columbia, 6224 Agricultural Road, Vancouver, B.C. V6T 1Z1, Canada}

\emailAdd{cbyrnes@cern.ch}
\emailAdd{lukas.hollenstein@unige.ch}
\emailAdd{rajeev.jain@unige.ch}
\emailAdd{urban@phas.ubc.ca}

\date{\today}

\abstract{We propose a novel scenario to generate primordial magnetic fields during inflation induced by an oscillating coupling of the electromagnetic field to the inflaton. This resonant mechanism has two key advantages over previous proposals. First of all, it generates a narrow band of magnetic fields at any required wavelength, thereby allaying the usual problem of a strongly blue spectrum and its associated backreaction. Secondly, it avoids the need for a strong coupling as the coupling is oscillating rather than growing or decaying exponentially. Despite these major advantages, we find that the backreaction is still far too large during inflation if the generated magnetic fields are required to have a strength of ${\cal O}(10^{-15}\, \Gauss)$ today on observationally interesting scales. We provide a more general no-go argument, proving that this problem will apply to any model in which the magnetic fields are generated on subhorizon scales and freeze after horizon crossing.}

\keywords{Inflation, primordial magnetic fields, backreaction}
\arxivnumber{1111.2030}

\begin{document}
\maketitle

%%%%%%%%%%%%%%%%%%%%%%%%%%%%%%%%%%%%%%%%%%%%%%%%%%%%%%%%%%%%%%%%%%%%%%%%%%%%%%%%%%%%%%
\section{Introduction}\label{GuessWhat}
%%%%%%%%%%%%%%%%%%%%%%%%%%%%%%%%%%%%%%%%%%%%%%%%%%%%%%%%%%%%%%%%%%%%%%%%%%%%%%%%%%%%%%

Large scale magnetic fields of $\mu\Gauss$ strength have been observed in cosmic structures at a great variety of length scales and redshifts: from stars to galaxies to regions around high redshift quasars as well as from clusters and superclusters to the region of low density filaments~\cite{Kronberg:1993vk,Han:2002ns,Govoni:2004as,Clarke:2000bz,Beck:2008ty,Guidetti:2007dd,Bonafede:2010xg,Xu:2005rb}.  Recently, different datasets have also been used to derive a lower bound of $B\gtrsim 10^{-15}\,\Gauss$ on the intensity of magnetic fields in the intergalactic medium~\cite{Neronov:1900zz,Tavecchio:2010mk,Tavecchio:2010ja,Dolag:2010ni,Ando:2010rb}. 
There exist various astrophysical mechanisms of magnetogenesis and the amplification of weak seed fields by adiabatic compression during structure formation and by the galactic dynamo mechanism seem well understood~\cite{Widrow:2002ud,Giovannini:2003yn,Brandenburg:2004jv,Brandenburg:1996fc}. Usually, primordial large scale magnetic fields are invoked to account for the seeds of nGauss strength needed for astrophysical amplification.  However, primordial magnetogenesis, active either during inflation or at phase transitions before recombination, is tightly constrained by Big-Bang Nucleosynthesis (BBN) and the Cosmic Microwave Background (CMB)~\cite{Durrer:2006pc,Paoletti:2010rx,Caprini:2011cw} (in addition, such models do generate non-gaussianities which can be used to constrain their model parameters~\cite{Seshadri:2009sy,Caprini:2009vk,Barnaby:2010vf,Trivedi:2011vt}), and moreover most mechanisms have difficulties generating the required field strength or the coherence length, see the reviews~\cite{Grasso:2000wj,Dolgov:2003xd,Kulsrud:2007an,Subramanian:2008tt,Subramanian:2009fu,Kandus:2010nw,Widrow:2011hs}. In fact, the required large correlation length is the most difficult property to explain. This applies equally well to astrophysical models and to primordial generation mechanisms.

One promising focus for the genesis of large scale magnetic fields is cosmological inflation.  Inflation is a quasi de Sitter stage of accelerated expansion of the Universe during which causally connected regions of space-time are stretched beyond the Hubble horizon. This enables the possibility of causal correlations on very large scales that re-enter the visible horizon after the end of inflation, either during the radiation or matter eras. In order for any amplification to occur during inflation it is necessary to step outside the Standard Model (SM) comfort zone because of conformal triviality of electromagnetism (EM). Since EM is conformally invariant and de Sitter space (as well as Friedmann-Lema\^itre-Robertson-Walker (FLRW) metrics in general) is conformally flat, this combination ensures that the Euler-Lagrange equations of motion (EOM) reduce to those in Minkowski space and no amplification, or any other non-trivial dynamics, can take place. Therefore one has to construct extremely efficient couplings in order to overcome this adiabatic behaviour. In principle this can be envisaged but in most scenarios the created EM energy density gives rise to issues with backreaction on the inflaton dynamics.

When a coupling of EM to the inflaton field (usually a scalar field) is implemented, it brings new terms in the EOM of the EM vector potential as well as the scalar field. For EM the coupling acts as an effective time-dependent mass term in the mode equation of the vector potential.  However, as it grows the energy density of the EM field can grow accordingly, up to the point where it destabilises the de Sitter background.  This can be shown to be a very general behaviour, and sets an upper limit on the maximal efficiency of a large class of mechanisms~\cite{Campanelli:2008kh,Demozzi:2009fu,Kanno:2009ei,Urban:2011bu}.  It turns out that in the most optimistic cases the final amplitude for the magnetic field, especially at large scales, is always several orders of magnitude below the desired level.  Similar considerations apply to axial-type couplings~\cite{Durrer:2010mq}.  The red thread which entwines these different mechanisms is the fact that all modes encompassed by inflation are amplified in a democratic way, from the lowest to the highest energies.  The broadness of the amplified EM spectrum then makes it extremely hard to safely allow for very intense magnetic fields at some scales, as the total energy density has to be shared with all modes. 

In order to circumvent this issue, we explore a novel avenue in which magnetic fields are resonantly amplified during inflation thanks to an axial coupling to the inflaton field\footnote{Earlier work focused on the parametric resonance of gauge fields during the oscillatory final stages of inflation (preheating) and found insufficient amplification in typical scenarios, see~\cite{Finelli:2000sh}.}.
Resonance happens in narrow bands such that only small ranges of modes can be amplified by this interaction. The resonance appears due to an oscillatory behaviour either of the coupling function itself or the dynamical evolution of the inflaton field\footnote{In general single field inflationary models with small periodic modulations in the potential~\cite{Chen:2008wn,Pahud:2008ae}, an interesting consequence of the oscillating behaviour of the scalar field is the resonant enhancements of the correlation functions of cosmological perturbations, see for instance Axion Monodromy Inflation~\cite{McAllister:2008hb,Flauger:2009ab}. It has been shown recently that such periodic modulations of the potential can lead to large resonant non-Gaussianity as well as other higher N-point functions~\cite{Flauger:2010ja,Chen:2010bka,Leblond:2010yq}.}.
The main purpose of this work is the analytic study of the consequences of such resonances and the requirements on the dynamics of the coupling. We demonstrate the feasibility of such a program by presenting a simple model at work. Despite the very different final EM spectrum, generation process and time, we will see that backreaction will still take place in a dangerous way before inflation can complete, and ultimately plague the final fate of this class of models.

We organise the presentation as follows. First, in Sec.~\ref{INFtoEM} we make the case for the coupling between inflation and EM, outlining the reasons behind our choice (the Mathieu equation) and study the structure of the mode equation for the EM potential.  Moving on to Sec.~\ref{Reconstruct} we will both generally and explicitly treat the issue of building a working model. 
Sec.~\ref{Generate} is where we present the analytic estimates for the generated EM spectrum. The issue of backreaction is discussed in Sec.~\ref{BackReactions} and we wrap things up in the concluding Sec.~\ref{TheEnd}. Appendix~\ref{conv} lists the conventions we use for describing the quantised EM vector potential and its spectrum during inflation. Finally, appendix~\ref{mathApp} contains a concise treatment of the Mathieu equation and its resonances.

%%%%%%%%%%%%%%%%%%%%%%%%%%%%%%%%%%%%%%%%%%%%%%%%%%%%%%%%%%%%%%%%%%%%%%%%%%%%%%%%%%%%%%
\section{An oscillating coupling between the inflaton and electromagnetism}\label{INFtoEM}
%%%%%%%%%%%%%%%%%%%%%%%%%%%%%%%%%%%%%%%%%%%%%%%%%%%%%%%%%%%%%%%%%%%%%%%%%%%%%%%%%%%%%%

%%%%%%%%%%%%%%%%%%%%%%%%%%%%%%%%%%%%%%%%%%%%%%%%%%%%%%%%%%%%%%%%%%%%%%%%%%%%%%%%%%%%%%
\subsection{Generalities and the Mathieu equation}

Our action is comprised of the free inflaton term $\cL_\cph$ and the EM Lagrangian density $\cL_{EM} = -F^2/4$, to which we add a chiral coupling to the inflaton
\be\label{intL}
  S_{\cph EM} \equiv \int\dd^4x\, \sqrt{-g}\, \frac{1}{4}f(\cph)\, F\tilde{F} \, ,
\ee
where $f(\cph)$ is a dimensionless coupling function and $\tilde{F}$ is the dual of the electromagnetic field tensor $F$, and $F\tilde{F}\equiv F_{\alpha\beta}\tilde{F}^{\alpha\beta}\equiv\frac{1}{2}\eta^{\alpha\beta\gamma\delta}F_{\alpha\beta}F_{\gamma\delta}$ with $\eta^{\alpha\beta\gamma\delta}$ being the totally antisymmetric Levi-Civita tensor with $\eta^{0123}=+|g|^{-1/2}$. This coupling introduces an interaction term of the form $f_\cph F\tilde{F}$ in the EOM of the scalar field (we use the notation $f_\cph\equiv\dd f/\dd\cph$).  However, as we investigate the amplification of electromagnetic fluctuations we will neglect this term in a first approximation, returning to it when discussing the backreactions on the dynamics of inflation.

We describe the electromagnetic field in terms of the vector potential in the Coulomb gauge (see appendix~\ref{conv} for our conventions), $(A_\mu)=(0,A_i)$ with $\partial_iA_i=0$. The classical evolution equation of its Fourier transform in the helicity basis, $\cA_\pm(k,\eta)$ reads
\be
  \cA''_h + \left(k^2 + h k f'\right) \cA_h = 0 \, ,
  \label{mode_eq}
\ee
where $h=\pm1$ stands for the positive and negative helicity, and $f'\equiv f_\cph\cph'$ is the conformal time derivative of the coupling function.  This equation remains identical when we analyse the evolution of quantum fluctuations, see for instance~\cite{Durrer:2010mq}. Note that one may also consider couplings of the form $I^2(\cph)\, F^2$ or $M(\cph)^2 A^2$ where $I(\cph)$ is a time-dependent coupling constant, and $M(\cph)$ is a dynamical photon mass, but we find it harder to resonantly generate magnetic fields in those cases.

The central idea of this work is to employ a radically different coupling structure, which discriminates among different energy scales (and selects only a small portion of them), and is efficient in the perturbative regime of~(\ref{intL}). These criteria can be met by an oscillating coupling term [c.f.\ Eq.~(\ref{g_eta})] which produces resonances.  In the specific case at hand, Eq.~(\ref{mode_eq}) possesses resonant solutions as long as the coupling term is a periodic function of our time variable.  We will focus on one specific case, the Mathieu equation, bearing in mind that any periodic coupling term would exhibit similar properties, and share qualitatively our results.  Such an oscillating term with a constant, or slowly time-varying frequency, automatically selects and amplifies only a very small range of Fourier modes, leaving all other modes almost unperturbed.  Furthermore, an EOM of the resonant type does not need any large parameter to produce appreciable enhancements.  Our choice is also guided by our aim to be able to provide as much analytic understanding as possible, in such a way to clearly pin down the advantages, shortcomings, differences and similarities compared to the previous literature on the subject.

\begin{figure}[t]
\includegraphics[width=0.48\textwidth]{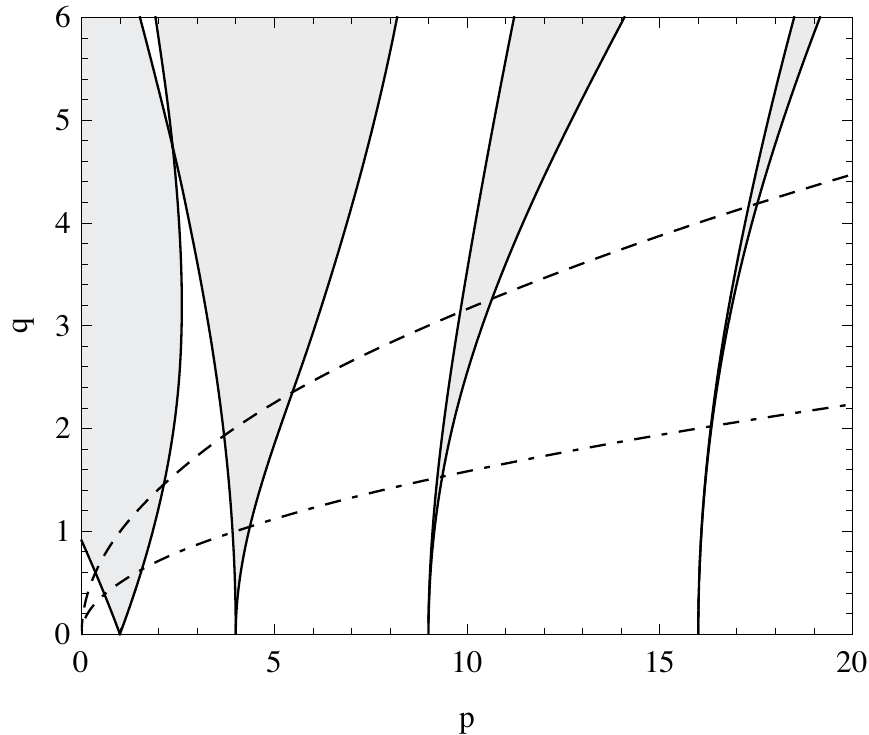}\hspace{3mm}
\includegraphics[width=0.48\textwidth]{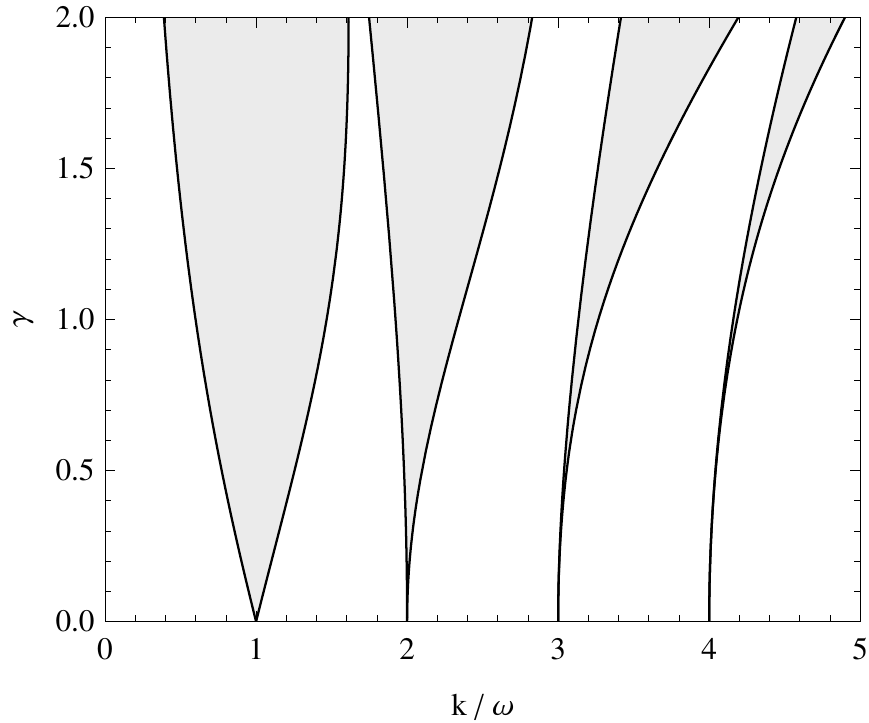}
\caption{\emph{Left panel:} The instability regions of the Mathieu functions are shaded in the $(p,q)$ plane. The solid lines show where the stability properties change. The broken lines correspond to $q=\sqrt{p}$ (dashed) and $q=\frac{1}{2}\sqrt{p}$ (dot-dashed). In terms of the dimensionless coupling constant,  $\gamma$, of the generic oscillating coupling function given in Eq.~(\ref{g_eta}) these represent the cases $\gamma=1,\,1/2$. \emph{Right panel:} The same chart but for our constrained parameters. Here $p=(k/\omega)^2$ and $q=\gamma k/\omega$.}
\label{fig:chart_aq}
\end{figure}

Let us swiftly review the main properties of the Mathieu equation (more details are available in appendix~\ref{mathApp}), which is typically written as
\be
  \frac{\dd^2 y(z)}{\dd z^2} + \left( p - 2q \cos 2z \right) y(z) = 0 \,.
  \label{mathieu}
\ee
Its solutions, the Mathieu functions, have well-studied instabilities in the $(p,q)$ plane where they exhibit exponential growth. We can define the evolution variable as $z\equiv\omega\eta$ for some constant frequency $\omega$ which is hence a free parameter of the model; therefore we can obtain a Mathieu equation [c.f.\ Eq.~(\ref{mode_eq})] by matching
\be
  p = k^2 / \omega^2 \, , \quad  2 q  \cos 2\omega\eta = k f' / \omega^2 \, ,
  \label{q_gprime}
\ee
up to the sign of the coupling function which only determines the phase of the oscillating part of the solution. This means that both helicity modes are amplified almost identically, up to a phase shift. The choice of the Mathieu equation therefore automatically fixes the coupling function $f(\eta)$ to be of the form
\be
  f(\eta) = \gamma \sin 2\omega\eta \, ,
  \label{g_eta}
\ee
where $\gamma$ is a dimensionless coupling constant that cannot depend on $k$ because $f(\cph)$ is only a function of the homogeneous background evolution, $\cph(\eta)$. Before discussing the coupling as a function of $\cph$, let us mention that in our specific case we work with a constrained $(p,q)$ set, since $q = \gamma\, k/\omega = \gamma \sqrt{p}$.  In Fig.~\ref{fig:chart_aq}, left, we show the stability chart of the solutions to the Mathieu equation in the $(p,q)$ plane. As examples, we draw the lines $q=\sqrt{p}$ and $q=\frac{1}{2}\sqrt{p}$ through the chart that represent the restricted parameter spaces for the cases $\gamma=1$ and $\gamma=1/2$. Also in Fig.~\ref{fig:chart_aq}, right, we use that restriction to draw the stability chart in the $(k/\omega,\gamma)$ plane where one can clearly see that the resonance bands are around $k=n\omega$ with integer $n$.

If a mode $\cA$ is resonantly amplified through the coupling to the inflaton, its amplitude roughly grows like $|\cA|\propto e^{\mu\omega\eta}$, where $\mu$ is the $k$-dependent characteristic exponent or Floquet index, and $\omega$ is the frequency of the driving term in the Mathieu equation (see appendix~\ref{mathApp} for details).  For now this frequency will be kept constant. As we will elucidate later, this choice is physically not very natural and thus we will generalise to a time-dependent $\omega$. Apart from the exponential growth the resonant solution also oscillates with constant amplitude. The effect of the opposite signs in front of the coupling functions for the two helicity modes of the vector potential, $\cA_\pm$, is a phase shift by $\pi/2$. As a consequence, the squared amplitudes of the two modes simply add up as
\be\label{matsol}
|\cA_+|^2 + |\cA_-|^2 \simeq |\cA_i|^2  e^{2\mu\omega\Delta\eta}
\,, \qquad \text{for}\ k \simeq n\omega, \ n\in\mathbb{N} \, .
\ee
Here $\Delta\eta\equiv\eta-\eta_i$ and $\cA_i\equiv\cA_\pm(\eta_i)$, assuming the same initial conditions for both helicities. The timespan $\Delta\eta$ for which a given mode is resonantly amplified is therefore sensitive to the time when the process starts which would generically be the beginning of inflation. Notice that herein lies the unnaturalness of this choice: the final result depends critically on the initial time where we want to be able to set generic initial conditions.  This will be alleviated when we allow for a time-dependent frequency which will also enlarge the range of modes that undergo resonant amplification.

%%%%%%%%%%%%%%%%%%%%%%%%%%%%%%%%%%%%%%%%%%%%%%%%%%%%%%%%%%%%%%%%%%%%%%%%%%%%%%
\subsection{Background evolution}\label{Reconstruct}

Before we turn to a careful examination of the solution~(\ref{matsol}) and its consequences, we need to understand how a Mathieu equation can arise in the context of (successful) inflationary models.  There are in general two ways to build a coupling function for which the dynamics of the inflaton result in the oscillatory behaviour that we need.

Recall that we chose to stick to a coupling to EM of the form $f(\eta) = \gamma \sin 2\omega\eta$. The simplest way to obtain the correct time evolution is to simply solve for the background dynamics, which returns the functional form for $\cph(\eta)$, and then engineer a coupling functional which translates this into the desired oscillations.  This path leads to a $f(\cph)$ which automatically yields the Mathieu resonances through the change of variables
\be\label{g_rec}
f(\cph) = \int \frac{\partial f(\eta)}{\partial \eta} \frac{\partial \eta}{\partial \cph} \dd\cph \, .
\ee
An obvious disadvantage of this approach is that the functional form of $f(\cph)$ may turn out to be quite unaesthetic. However, the point we want to make here is not one of attractiveness, but to show how in principle it is possible to build a working model of inflation which resonantly couples to EM.

As an explicit example, we construct the coupling in the case of chaotic inflation with the potential $V(\cph)=\frac{1}{2}m^2 \varphi^2$ and the slow roll solution $\varphi^2\simeq4\, \mpl^2 N$, where $N$ is the number of e-folds until the end of inflation at $\eta_f$. A coupling of the form
\be\label{coupetaout}
f(\cph) = \gamma \sin\!\left[ 2\,\omega\,\eta_f\, e^{(\cph/2\mpl)^2-1/2}\right],
\ee
leads to the Mathieu equation and therefore the EM modes within the resonance bands are amplified exponentially as long as the slow roll approximation holds.

Another solvable model is hybrid inflation with the potential $V(\cph)=V_0 (1+ \frac{1}{2}\frac{m^2}{\mpl^4}\cph^2)$ where $V_0$ sets the energy scale of inflation and $m^2\cph^2 \ll \mpl^4$ for all values traversed by the inflaton.  Inflation ends when the waterfall field becomes destabilised at $\cph=\cph_f$, which may be chosen arbitrarily.  Since $\varphi(N)\simeq \varphi_f \exp(N\,m^2/\mpl^2)$, we find the required coupling to be
\be
f(\cph) = \gamma \sin\left[2\,\omega\,\eta_f \left(\varphi/\varphi_f\right)^{\mpl^2/m^2}\right] \, .
\ee

The second possibility is to choose a simple coupling functional $f(\cph)$ and work out the inflaton Lagrangian implied by $\cph(\eta)$ that is known from the chosen form of $f(\eta)$. Despite the more attractive interaction term this simply transfers the complication to the form of $\cL_\cph$.  However, the inflaton field must roll {\it down} to the minimum of the potential and its field value must be a monotonic function of time. Similarly $\dot{\cph}$ must always keep the same sign until reheating after the end of inflation. Hence both of these quantities can only contain a subdominant oscillating part, which is likely to make it difficult to construct an efficient coupling. On the other hand, the second derivative $\ddot{\cph}$ could oscillate so a coupling of the form $f(\Box \cph)$ might work more efficiently, even though it would be necessarily suppressed by two more powers of the mass scale associated with the interaction. Due to these issues we focus on the first approach in this work.

%%%%%%%%%%%%%%%%%%%%%%%%%%%%%%%%%%%%%%%%%%%%%%%%%%%%%%%%%%%%%%%%%%%%%%%%%%%%%%%%%%%%%%
\section{Resonant magnetic field generation}\label{Generate}
%%%%%%%%%%%%%%%%%%%%%%%%%%%%%%%%%%%%%%%%%%%%%%%%%%%%%%%%%%%%%%%%%%%%%%%%%%%%%%%%%%%%%%

With the background dynamics now specified, we are in a position to look at the evolution of the EM potential in some detail.  In this section we first make analytic estimates of the resonant amplification in the simple case of a constant oscillation frequency. We then move on to the  case of a time-dependent frequency which turns out to be more physical.

%%%%%%%%%%%%%%%%%%%%%%%%%%%%%%%%%%%%%%%%%%%%%%%%%%%%%%%%%%%%%%%%%%%%%%%%%%%%%%%%%%%%%%
\subsection{Constant frequency}

If the frequency $\omega$ of the coupling is constant, only the Fourier modes within the resonance bands given by $\gamma$ and $\omega$ are amplified. In order to ensure that we are working in the perturbative regime of EM we take the coupling constant to be small, $\gamma\ll 1$. Looking at Fig.~\ref{fig:chart_aq} we can see that for $\gamma\ll1$ the resonance bands are very narrow in $k/\omega$. The first band is located at $k=\omega$ and is by far the most important and efficient. Its width can be estimated to be $\simeq \gamma$ such that its boundaries are roughly at
\be\label{kpm}
k_\pm \equiv \left( 1 \pm \frac{\gamma}{2} \right) \omega \, .
\ee
Within this first band the Floquet index can be estimated as
\be\label{floq1}
  \mu_k \simeq \left[ \gamma^2/4 - (k/\omega-1)^2 \right]^{1/2}
\ee
such that we have $\mu_{k=\omega}\simeq \gamma/2$ at the peak. We only work with the first resonance band for the analytic estimates, as most of the EM energy density will be stored in this narrow band of modes.

A resonant mode $k \in [k_-,k_+]$ begins to grow exponentially according to (\ref{matsol}) at the time inflation starts, $\eta_i$. The amplification lasts until $k$ crosses the horizon at $\eta_k \equiv -1/k$, because afterwards the cosine in the EOM does not go through any further period anymore and the oscillations of the coupling are effectively turned off. Hence, the time for which solution~(\ref{matsol}) describes the evolution of the mode $k$ well is estimated as
\be\label{deltaeta}
\Delta\eta = \eta_k - \eta_i \,.
\ee
Now we can estimate the power spectrum of magnetic fields amplified by the resonance. The power spectrum is $\delta_B^2 \propto k^5\sum_h |\cA_h|^2$, see appendix~\ref{conv}, and the squared amplitudes of the resonant modes grow as $\exp(2\mu\omega\Delta\eta)$ according to Eq.~(\ref{matsol}). Since the resonance band is so narrow we can take the peak value for the Floquet index, $\mu\simeq\gamma/2$, and model the $k$-dependence by introducing a narrow dimensionless window function $W(k/\omega,\gamma)$ centred at $k=\omega$ that drops to 0 beyond $k_\pm$. Finally, we neglect $\eta_k$ with respect to $\eta_i$ in $\Delta\eta$ which is a good approximation for all modes that exit the horizon not too shortly after the beginning of inflation, as is the case even for modes which stretch up to 1 Gpc today. With these approximations we obtain the following spectrum at the end of inflation
\be\label{spike}
\delta_B^2 = \frac{k^5 |\cA_i(k)|^2}{(2\pi)^2 a_f^4} \, W(k/\omega,\gamma)\, \exp(-\gamma\,\omega\,\eta_f\, e^{N_f}) \, .
\ee
after subtraction of the vacuum outside the amplified resonance band, $[k_-,k_+]$. Here $a_f=a(\eta_f)$ and $\eta_i = \eta_f e^{N_f}$ with the total number of e-folds $N_f$ and conformal time at the end of inflation $\eta_f$. The initial amplitude of the mode function, $|\cA_i(k)|^2$, is given by the vacuum normalisation and has units $[1/k]$.

The total EM energy density is given by $\int \dd\ln\!{}k\,(\delta_B^2+\delta_E^2)$ where the electric field spectrum is $\delta_E^2\propto k^5|\cA_h'|^2$. However, estimating the conformal time derivative of the resonant modes gives
\be\label{eenergy}
|\cA_h'|^2 \simeq \mu^2 \omega^2 |\cA_h|^2 \ll k^2 |\cA_h|^2 \, ,
\ee
and the contribution from the electric fields is negligible compared to that of the magnetic fields. For simplicity let us assume a standard Bunch-Davies vacuum initially, $|\cA_i|^2=(2k)^{-1}$, and replace the window function with a top-hat filter. Doing so we can integrate the spectrum to find the energy density at the end of inflation
\be\label{resultk}
\rho_{EM}^f \simeq \gamma \left( 1 + \frac{\gamma^2}{4} \right) \frac{\omega^4}{8\pi^2 a_f^4} \exp(-\gamma\,\omega\,\eta_f\, e^{N_f})\, .
\ee
However, we will see in the next section that backreaction will spoil inflation if this quantity is non-negligible.

A very unattractive feature of this mechanism is the explicit dependence of the spectrum and the energy density on the initial time $\eta_i$, and therefore the double-exponential dependence on the total number of e-folds $N_f$. The reason for this is of course the choice of a constant frequency $\omega$, which operates irrespective of the background dynamics.  In addition to that, this frequency refers to periods in terms of the $\eta$ conformal clock, which is itself exponentially changing with $N$. In the next section we shall see how introducing a time-dependent frequency can mitigate this problem and give a physically more natural behaviour.

%%%%%%%%%%%%%%%%%%%%%%%%%%%%%%%%%%%%%%%%%%%%%%%%%%%%%%%%%%%%%%%%%%%%%%%%%%%%%%%%%%%%%%
\subsection{Time-varying frequency}

We have seen that although a constant frequency does resonantly produce magnetic fields, the final result turns out to be sensitive to the total number of e-folds of inflation in a doubly exponential manner. In fact any direct sensitivity to the beginning of inflation feels unnatural and would bring about a dependence on the initial conditions of inflation, which is the opposite of what inflation is ordinarily invoked for.  This may also lead to problems with the initial quantisation of the fields.  The most obvious amelioration to the simplest case discussed above is to allow for a time-dependence in $\omega$.  Provided the time variation of the frequency is small  compared to the oscillation time scale, the Mathieu equation will still be a good approximation to the resulting behaviour. 

In order to be able to set the initial conditions for all the modes in a consistent way, we choose the frequency
\be\label{omegaeta}
\omega(\eta) \equiv \hat\omega\, e^{\beta\eta} \, ,
\ee
where the strength of the time-variation is tuned with $\beta$ while $\hat\omega$ is the pivotal resonant frequency.  In this case the resonance will not be operational throughout inflation for the same mode $k$, but will move across a range of modes.  Thus, in addition to curing the initial condition problem, this allows for a more controlled amplification of a wider range of modes. This clearly makes it a more physical model.

\begin{figure}[t]
\centering
\includegraphics[width=\textwidth]{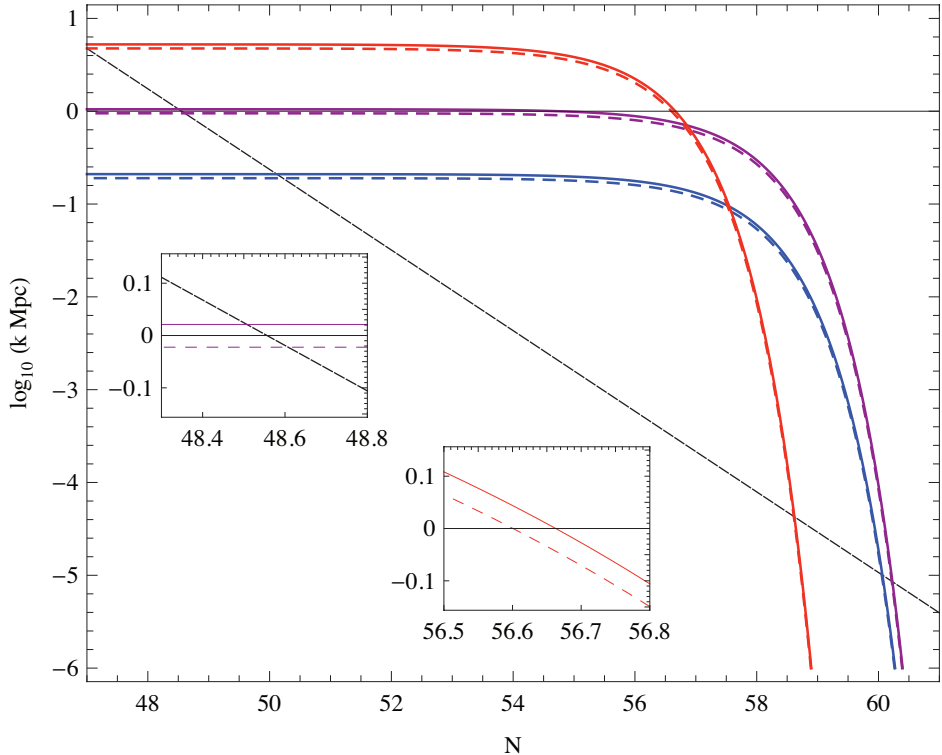}
\caption{The number of e-folds away from the end of inflation at which a mode $k$ enters and leaves the resonance band, $N(\eta_+(k))$ solid and $N(\eta_-(k))$ dashed, are given for the model $\omega=\hat\omega e^{\beta\eta}$. The blue curves correspond to $\hat\omega=0.2/\Mpc$, $\beta=10^{-4}/\Mpc$; the purple curves to $\hat\omega=1/\Mpc$, $\beta=10^{-4}/\Mpc$; and the red curves are for $\hat\omega=5/\Mpc$, $\beta=10^{-3}/\Mpc$. We set $\gamma=0.1$ for all curves. The black dashed line is $N(\eta=-1/k)$ where the given mode exits the horizon scale.  We zoom on the evolution of the mode $k=1/\Mpc$ (solid black) for the second (purple curve) and third (red curve) choice of parameters: in the former case the mode becomes superhorizon before leaving the resonance band; the latter instead sees the resonance band fully traversed within the Hubble scale.}
\label{fig:Npm}
\end{figure}

Given the simplicity of this model, we can make a number of reasonable approximations which will enable us to understand the resulting spectrum in an analytic way. The first step is to identify which modes resonate, and for how long. A mode $k$ enters ($+$) and leaves ($-$) the resonance band at $\eta_\pm$ given by $k=(1\pm\gamma/2)\omega(\eta_\pm)$, c.f.~Eq.~(\ref{kpm}). We find
\be\label{whenres}
\eta_\pm = \frac{1}{\beta} \ln\left[ \frac{k}{(1\pm\gamma/2)\,\hat\omega} \right] \,.
\ee
We show these relevant times for three examples in Fig.~\ref{fig:Npm}. In this case a mode is resonating for
\be\label{timespan}
\Delta\eta = \eta_- -\eta_+
= \frac{1}{\beta} \ln \left[\frac{1+\gamma/2}{1-\gamma/2}\right]
\simeq  \frac{\gamma}{\beta} \,,
\ee
where we used $\gamma\ll 1$. Only for $k\sim\hat\omega$ do we need to worry about $k$ exiting the horizon before leaving the resonance band. In this case $\Delta\eta=\eta_k-\eta_+\simeq \gamma/2\beta$, where we can safely ignore $\eta_k$ because typically $\beta\ll\hat\omega$. We observe that only the modes with $\beta \lesssim k \lesssim \hat\omega$ pass through the resonance before horizon exit. Notice that the time spent in resonance band neither depends on $k$ nor on the number of e-folds, a fact that reflects the more physical nature of the model with time-dependent frequency.

With these estimates it is immediate to compute the amplification factor experienced by a resonant mode. The crudest approximation is obtained assuming that the Floquet index is a constant across the narrow resonance, $\mu_k\simeq\gamma/2$, and taking $\omega=k$ at the time of resonance we find
\be\label{growth}
|\cA_+|^2+|\cA_-|^2 \simeq |\cA_i|^2\, e^{\gamma^2 k/2\beta} \, 
\ee
(the factor of $1/2$ in the exponent accounts for the fact that for $k\sim\hat\omega$ the resonance is active for only about half of~(\ref{timespan})). Now that the coupling term in the mode equation is negligible well inside the horizon, we can safely set the Bunch-Davies initial conditions such that the spectrum is $\delta_B^2 \propto k^4 \exp \left[\gamma^2 k/2\beta\right]$.  A much better approximation can be found if we keep the full expression~(\ref{floq1}) for $\mu_k$ and integrate the exponent over time, which can still be done analytically, to find
\be\label{spectrumPs}
\delta_B^2 = \frac{k^4}{(2\pi)^2 a_f^4} \left\{ \frac{1}{2} \exp\left[ \pi \gamma^2 k / 8 \beta \right]
\theta(k - \beta)\theta(\hat\omega - k) + \theta(k - \hat\omega) \right\} \, ,
\ee
where we have included the unamplified part for completeness, and the $\theta$ are Heaviside unit step functions. This expression is quite accurate for $\hat\omega/\beta \lesssim 10^4$ or so.  The spectrum is very blue in the amplified region, $k\in [\beta,\hat\omega]$. The two parameters of the model, $\beta$ and $\hat\omega$, serve as natural IR and UV cutoffs in the theory.  In order to have some amplification, $\beta$ should be small compared to the pivotal mode we want to amplify, $\hat\omega$.
Let us finally compute the EM energy density by integrating the spectrum.
\be\label{energyeta}
\rho_{EM}^f \simeq \frac{\beta \hat\omega^3}{\pi^3 \gamma^2 a_f^4}\, 
  \exp\left[\pi\gamma^2\hat\omega / 8\beta \right] \, .
\ee
Similarly to the constant $\omega$ case we can safely neglect the electric field energy density and, therefore, the contribution to the total EM energy density comes only from the magnetic fields.

%%%%%%%%%%%%%%%%%%%%%%%%%%%%%%%%%%%%%%%%%%%%%%%%%%%%%%%%%%%%%%%%%%%%%%%%%%%%%%%%%%%%%%
\section{Backreactions and subhorizon resonances}\label{BackReactions}
%%%%%%%%%%%%%%%%%%%%%%%%%%%%%%%%%%%%%%%%%%%%%%%%%%%%%%%%%%%%%%%%%%%%%%%%%%%%%%%%%%%%%%

The strong feedback from the generated EM energy density on the dynamics of the inflaton is a problem for many inflationary mechanisms of magnetogenesis. In our model there are two key properties that set it apart from most other scenarios: first, the resonant amplification is operational while a given mode is within the Hubble scale and freezes outside. Second, only a very narrow band of modes is amplified and not an entire power law spectrum. These two characteristics compete for the final outcome of the backreaction calculation. A spectrum that is a narrow window at large scales reduces the total EM energy density as compared to that of a power law spectrum. On the other hand, the fact that the process ceases after horizon crossing demands a very large amplification factor in the first place. This is because the EM energy density dilutes like $a^{-4}$ all the way from horizon exit until the end of inflation. We therefore need a huge magnetic field at horizon crossing in order for it to have an appreciable strength today. The larger the scale we consider, the earlier it exits and the more severe the problem becomes.

Trying to balance these two effects, we find that backreaction always sets in too early and inflation is severely compromised.  For instance, we can follow the growing magnetic field until it eventually catches up with the inflaton energy density. Or we can ask how long inflation could possibly last if we were to generate, at a given scale, a magnetic field to match the observed ones. Taking the second path, it is easy to have a rough idea of which parameter values would be needed in order to achieve a magnetic field that today peaks at 1 nGauss at 1 Mpc.  Working with the time-dependent frequency~(\ref{omegaeta}), and fixing $\hat\omega = 1/\Mpc$ we find that the desired intensity
\be\label{desired}
\delta_B^0(k=\hat\omega) = \frac{\hat\omega^2}{2\sqrt2\pi} e^{\pi \gamma^2 \hat\omega / 16\beta} = 1\,\text{nGauss}
\quad\Rar\quad \beta \approx 1.75\times10^{-3} \gamma^2 \,\Mpc^{-1} \, .
\ee
We can now plug~(\ref{desired}) into~(\ref{energyeta}) to find that the EM energy density at the end of inflation is about $10^{-9}\rho_\cph$ (where, for simplicty, $\rho_\cph \simeq V(\cph)$).  Hence, it may appear that the generated magnetic fields are safely subdominant and inflation can proceed unharmed.  However, in order to obtain these values at the end of inflation, we have to trace back the EM energy density to the point where the amplified window of modes left the horizon by taking into account the superhorizon dilution of $1/a^4$; as we shall show in the next section, contrary to our expectations, the EM energy density greatly overtakes that of the inflaton at horizon crossing.

%%%%%%%%%%%%%%%%%%%%%%%%%%%%%%%%%%%%%%%%%%%%%%%%%%%%%%%%%%%%%%%%%%%%%%%%%%%%%%%%%%%%%%
\subsection{No-Go for subhorizon amplification}\label{nogo}

In conclusion, it is hard to generate cosmological magnetic fields at large scales strong enough to be measured today, or even to provide seeds for some late-time amplification mechanisms.  This seems to be a manifestation of a more general pattern which is encountered in coping with subhorizon amplification mechanisms. Amplification of modes within (or around) the horizon and freezing outside appears to be a property of models with generic axial couplings even without resonances, the reason being that the coupling is suppressed by a factor $k$ outside the horizon, see \cite{Durrer:2010mq}.

We will present a more formal ``proof'' of this fact here.  The assumptions, encompassing the models we have analysed so far, are:
\begin{enumerate}
	\item the amplification mechanism is only active inside the horizon, and ceases, with consequent freezing of the vector potential, once a mode crosses the horizon;
	\item the only effect of the reheating stage following inflation is to alter the overall decay of the total inflaton energy density and the evolution of the scale factor, and has no direct impact on the EM field.
\end{enumerate}
Let us start with the aim of producing a magnetic field strength $B_{0*}$ today at a correlation scale $k_*$. To minimise backreaction we minimise the total magnetic energy density by considering a power spectrum that is a Dirac delta peak, $P_S=A k_*\delta_{\cal D}(k/k_*-1)$.  Then the energy density is $\rho^0_{B}=Ak_*^4/(2\pi)^2$ and the coefficient $A$ is chosen such that $\rho^0_{B}=\frac{1}{2}B_{0*}^2$.  We also neglect the energy density in the electric field during inflation, in order to be as conservative as possible. We can then calculate the energy density of the magnetic field $N$ e-folds before the end of inflation
\be\label{e4nEvol}\label{youwish}
\rho_B(N) = \rho_B^{end} e^{4N} =\rho_B^0 a_{end}^{-4}\, e^{4N}  \, ,
\ee
where $a_{end}$ is the scale factor at the end of inflation (normalised today) and $\rho_B^{end}$ is the magnetic energy density at the end of inflation. We have assumed that the $a^{-4}$ behaviour is obeyed throughout the evolution of the post-inflationary Universe as well.\footnote{This assumption is well justified since we are looking at very large scales, and there is no risk of incurring any (direct or inverse) cascades. Even if an inverse cascade was active it is unlikely that it could have a huge effect, see e.g.~\cite{Durrer:2010mq}.}

The backreaction problem is caused by the fact that $\rho_B$ decays much more quickly than the energy density of the inflaton. The simplest and most optimistic way to treat the evolution of the Universe after inflation is to assume it was always radiation dominated.  In that case the ratio $\Omega_B \equiv \rho_B/\rho_{\rm total}$ is constant after inflation ends.  For a conservative magnetic field strength of $B_{0*}=10^{-15}\, \Gauss$ today we have $\Omega_B\approx 10^{-23} \ll 1$, and for a more realistic calculation it would clearly be larger.
However even for this optimistic case, we can see that, evolving the magnetic field backwards into inflation, its energy density will dominate after only
\be
N_* = -\frac{1}{4}\ln \Omega_B \approx 13
\ee
e-folds if we assume a de-Sitter expansion, independent of the scale $k_*$ and the inflationary energy scale. More realistically, in slow roll the inflaton energy density slightly increases going backwards into inflation which improves the situation marginally. However, the observational bound on the tensor-to-scalar ratio means that the energy density cannot increase very significantly. This clearly demonstrates that we could only have magnetic fields with an interesting amplitude ($B_{0*} \gtrsim 10^{-15}$ Gauss) on very small scales, $k\geq k_*=\cH(N_*)$, corresponding to scales which crossed the horizon as late as $N_* \lesssim 13$ e-folds before the end of inflation. The problem is alleviated a little with a lower energy scale for inflation, as large scales would exit the horizon at smaller $N$, but the improvement is minimal.

One may turn this reasoning around, and ask what is the maximum magnetic field strength which could be produced within the limits set by the backreaction.  In this case we can fix the ratio $\Omega_B(N)=1$ at the time of horizon exit, which is the earliest time at which we have absolute confidence in the time-evolution of the magnetic field (without specifying a concrete model).  From then on and until the end of inflation this ratio dilutes away as $e^{-4N}$: the final ratio, for large scale modes which typically cross the Hubble scale around $N \gtrsim 50$ is therefore $\Omega_B\ll10^{-87}$, and the corresponding magnetic field today would be $B_0 \ll 10^{-47}$ Gauss.  Hence, any reasonable duration for inflation automatically leaves us with only crumbs of magnetic energy today.

Finally, we should emphasise that this no-go theorem is not specific to the case of a narrowly peaked magnetic power spectrum. As mentioned above, assuming a delta function for the spectrum simply is the most conservative choice and any broader distribution or power law spectrum gives even more stringent constraints.

%%%%%%%%%%%%%%%%%%%%%%%%%%%%%%%%%%%%%%%%%%%%%%%%%%%%%%%%%%%%%%%%%%%%%%%%%%%%%%%%%%%%%%
\section{Summary and conclusions}\label{TheEnd}
%%%%%%%%%%%%%%%%%%%%%%%%%%%%%%%%%%%%%%%%%%%%%%%%%%%%%%%%%%%%%%%%%%%%%%%%%%%%%%%%%%%%%%

In this work, we have made a novel proposal for the generation of primordial magnetic fields induced by an oscillating axial interaction between the inflaton and the EM field. This may lead to a highly efficient resonant production of magnetic fields during inflation, and, unlike other scenarios, without the need for a coupling whose value grows or decays dramatically during inflation.  Through studying the Mathieu equation, we can see that only certain wavelengths, which correspond to instability bands leading to exponential growth, will be amplified, which is another key advantage of our scenario.  While most alternative scenarios ultimately lead to a blue magnetic field spectrum because of the onset of the redder and stronger electric field, in our scenario one may tune the parameters of the coupling so that only modes of observational interest are amplified, leaving the electric field modes behind; this leads to a sharply peaked spectrum, which will spread over a more realistic range of wavelengths by the further evolution in the plasma.

We have proven how for any inflationary potential one may construct a coupling such that the magnetic fields experience the resonances of the Mathieu equation, both through a general formula for reconstructing the coupling and through two explicit examples. The results are rather complicated. However, we only chose to work with the exact Mathieu to be able to provide analytic estimates of the generated fields and to neatly capture the underlying physics. Notice that any periodic coupling would give rise to resonances and similar spectra and we therefore do not have to restrict to exactly those couplings derived in Sec.~\ref{Reconstruct}. Alternatively, and even in the case of the Mathieu equation, one may choose a simple coupling and instead  engineer the dynamics of the inflaton to generate the same resonance.

We have studied both the case of a constant resonant frequency, which is analytically simplest but sensitive to the beginning of inflation, and the more realistic case of a slowly varying coupling.  In the case of a constant frequency, the modes in a resonance band will grow exponentially from the beginning of inflation until horizon crossing, leading to a doubly exponential sensitivity (in terms of the number of e-folds) on the initial time of inflation, and to theoretical embarrassment in finding a consistent set of initial conditions.  We have shown how a slowly varying frequency solves these problems.

Generating primordial magnetic fields with large correlation scale and without backreacting is notoriously difficult.  We have outlined above several reasons why our new method should offer significant advantages.  Unfortunately the backreaction does remain a severe problem in our scenario, even when the magnetic field is concentrated in a very narrow band and negligible everywhere else.  The issue is that the magnetic field generation only takes place while the chosen modes are still subhorizon, and after horizon crossing they freeze.  Hence, their energy density is diluted like $a^{-4}$ after crossing the horizon. This is not so severe after inflation until today, since the radiation and matter densities also decay rapidly. However, the effect during inflation is dramatic because the inflaton energy density is nearly constant by construction, and so the relative energy density of the magnetic field will be diluted exponentially in terms of the number of e-folds from horizon crossing until the end of inflation.

Under the most optimistic assumptions, we have traced a single Fourier mode of strength $10^{-15}\, \Gauss$ today backwards into inflation. We found that already $N=13$ e-folds before the end of inflation backreaction sets in, compromising the course of slow roll inflation. It is therefore clear that our mechanism, despite its advantages, cannot be invoked to generate large scale magnetic fields of sizeable strength today.

We emphasise that this no-go result does not only hold for the mechanism studied here. It can directly be applied to all mechanism of inflationary magnetogenesis where the vector potential modes are constant outside the horizon. Only special reheating scenarios or inverse cascades in the plasma era that change the shape and/or amplitude of the magnetic spectrum in an extreme way could potentially mitigate this result.  In conclusion, the problem of the origin of cosmic magnetic fields again remains wide open.

\indent {\bf Note added.} After this work appeared on the arxiv, a quite general no-go result for primordial magnetogenesis was presented in~\cite{Bonvin:2011dr,Bonvin:2011dt}; in which it was shown that the magnetic fields generated during inflation typically yield unacceptably large anisotropic stresses during the subsequent radiation era. It has not been checked whether this result will apply to the model proposed in this paper.

\acknowledgments
The authors thank Ruth Durrer for useful discussions. CTB and FU thank the D\'epartement de Physique Th\'eorique at Universit\'e de Gen\`eve for kind hospitality.  LH thanks Bielefeld University for hospitality.  LH and RKJ acknowledge financial support from the Swiss National Science Foundation.

%%%%%%%%%%%%%%%%%%%%%%%%%%%%%%%%%%%%%%%%%%%%%%%%%%%%%%%%%%%%%%%%%%%%%%%%%%%%%%%%%%%%%%
\appendix

%%%%%%%%%%%%%%%%%%%%%%%%%%%%%%%%%%%%%%%%%%%%%%%%%%%%%%%%%%%%%%%%%%%%%%%%%%%%%%%%%%%%%%
\section{Conventions}\label{conv}
%%%%%%%%%%%%%%%%%%%%%%%%%%%%%%%%%%%%%%%%%%%%%%%%%%%%%%%%%%%%%%%%%%%%%%%%%%%%%%%%%%%%%%

We work with a metric signature \mbox{($-$ + + +)}.  For tensor components, Greek indices take values $0\ldots3$, while Latin indices run from $1$ to $3$.  We employ Heaviside-Lorentz units such that $c=\hbar=k_B=\epsilon_0=\mu_0=1$. The reduced Planck mass is defined as $\mpl=(8\pi G)^{-1/2}$. We normalise the cosmic scale factor to unity today so that comoving scales become physical scales today.

For the EM fields, our conventions follow~\cite{Barrow:2006ch}:
\be
  E_\mu = F_{\mu\alpha} u^\alpha \, , \qquad
  B_\mu = \frac{1}{2}\eta_{\mu\alpha\beta\gamma} F^{\alpha\beta} u^\gamma
   = \tilde F_{\mu\alpha} u^\alpha \,,
\ee
where $E_\alpha u^\alpha=0=B_\alpha u^\alpha$.  In a perturbed FLRW metric an observer has the four-velocity $u^\mu=a^{-1}(1,\V{0})+{\cal O}(1)$. As a consequence one finds (in Coulomb gauge)
\be
  \left(E_\mu\right) = \Big(0,\ -\frac{1}{a} A'_i \Big) \, , \qquad
  \left(B_\mu\right) = \Big(0,\ \frac{1}{a}\epsilon_{ijk}\partial_j A_k \Big) \, .
\ee
Introducing the orthonormal spatial basis
\be
  \left(\Veps^\Vk_{1}\,,\,\Veps^\Vk_2\,,\,\hat\Vk\right) \qquad
  \text{with} \quad  |\Veps^\Vk_{i}|^2=1 \,,\ \hat\Vk=\Vk/k \,,
\ee
and the helicity basis
\be
  \Veps^\Vk_{\pm} \equiv \frac{1}{\sqrt{2}}\left( \Veps^\Vk_1 \pm i\Veps^\Vk_2 \right) \,,
\ee
the vector potential takes the form $\cA_+\Veps_++\cA_-\Veps_-$. After quantisation of the vector potential, we can study the evolution of the Fourier modes, $\cA_h(\eta,k)$, with respect to the helicity basis for the polarisation states, $h=\pm$.

If the magnetic field generated by some process is statistically homogeneous and isotropic, its spectrum is determined by two scalar functions $P_S(k)$ and $P_A(k)$. Since the magnetic field is a divergence-free vector field the two-point function of the Fourier components of the magnetic field can be written, for a comoving wave vector $\Vk$, as
\be
  \langle \widetilde{B}_i(\eta,\Vk) \widetilde{B}^*_j(\eta,\Vq) \rangle
  \ =\ \frac{(2\pi)^3}{2} \delta(\Vk-\Vq)
  \Big\{(\delta_{ij} -\hat{k}_i\hat{k}_j) P_S(\eta,k) - i \epsilon_{ijn} \hat{k}_n P_A(\eta,k) \Big\} \, ,
\ee
where $P_S$ and $P_A$ are the symmetric and anti-symmetric parts of the power spectrum, respectively. The symmetric part of the spectrum determines the energy density:
\be
  \langle \widetilde{B}_i(\eta,\Vk) \widetilde{B}^*_i(\eta,\Vq) \rangle
  \ =\  (2\pi)^3\delta(\Vk-\Vq)P_S(\eta,k) \, .
\ee
With respect to the helicity basis the spectra can be directly written as
\be\label{e:PSA}
  P_{S/A}(\eta,k)\ =\ k^2\left( |\cA_+(\eta,k)|^2 \pm |\cA_-(\eta,k)|^2 \right) \,,
\ee
where the upper sign corresponds to $P_S$ and the lower sign to $P_A$. Here we use the non-trivial result widely applied in inflationary cosmology that at late times, the vacuum expectation values of the fields generated during inflation can be interpreted as stochastic power spectra. Finally, sometimes in the literature the power spectrum is identified with
\be\label{appDeltaB}
\tilde\delta_B^2 \equiv \frac{k^3}{(2\pi)^2} P_S(\eta,k) \quad\Rar\quad \delta_B^2 = \frac{k^3}{(2\pi)^2 a^4} P_S(\eta,k) \,,
\ee
which is often denoted as $\dd\rho_B/\dd\ln k=\delta_B^2$. When integrated
\be
\rho_B=\int_0^\infty\frac{\dd k}{k}\, \delta_B^2 \,,
\ee
it gives the average magnetic energy density, $\rho_B=\frac{1}{2}B^2$.

%%%%%%%%%%%%%%%%%%%%%%%%%%%%%%%%%%%%%%%%%%%%%%%%%%%%%%%%%%%%%%%%%%%%%%%%%%%%%%%%%%%%%%
\section{The Mathieu equation, stability and characteristic exponents}\label{mathApp}
%%%%%%%%%%%%%%%%%%%%%%%%%%%%%%%%%%%%%%%%%%%%%%%%%%%%%%%%%%%%%%%%%%%%%%%%%%%%%%%%%%%%%%

The Mathieu equation we are working with is written in the form~(\ref{mathieu}):
\be
  \frac{\dd^2 y(z)}{\dd z^2} + \left( p - 2q \cos 2z \right) y(z) = 0 \,.
  \label{mathieuagain}
\ee

\paragraph{Stability chart:}
In order to find in which regions of the $(p,q)$ plane (recall that $p$ is essentially a normalised $k^2$) the solutions to the differential equation diverge, we need to identify the solutions with period $\pi$ or $2\pi$.  It turns out that such solutions are always expressible in series of sines or cosines such as
\beq
\text{Period}\;\pi && \left\{
\begin{array}{l}
  y(z) = \sum_{n=0}^\infty a_n\cos 2nz \\
  y(z) = \sum_{n=0}^\infty b_n\sin 2nz
\end{array}\right. \, ,\\
\text{Period}\;2\pi && \left\{
\begin{array}{l}
  y(z) = \sum_{n=0}^\infty c_n\cos (2n+1)z \\
  y(z) = \sum_{n=0}^\infty d_n\sin (2n+1)z
\end{array}\right. \, .
\eeq
Plugging these back into the original Mathieu equation, we find a system of recursive relations such as
\beq
\text{System A} && \left\{
\begin{array}{l}
  p a_0 - q a_1 = 0 \\
  (p-4)a_1 - q(2a_0 + a_2) = 0 \\
  (p-4n^2)a_n - q(a_{n-1} + a_{n+1}) = 0 \quad (n \geq 2)
\end{array}\right.\label{sysa} \, ,\\
\text{System B} && \left\{
\begin{array}{l}
  (p-4)b_1 - q b_2 = 0 \\
  (p-4n^2)b_n - q(b_{n-1} + b_{n+1}) = 0 \quad (n \geq 2)
\end{array}\right.\label{sysb} \, ,\\
\text{System C} && \left\{
\begin{array}{l}
  (1-p+q)c_0 + q c_1 = 0 \\
  \left[(2n+1)^2 - p\right]c_n + q(c_{n-1} + c_{n+1}) = 0 \quad (n \geq 1)
\end{array}\right.\label{sysc} \, ,\\
\text{System D} && \left\{
\begin{array}{l}
  (1-p-q)d_0 + q d_1 = 0 \\
  \left[(2n+1)^2 - p\right]d_n + q(d_{n-1} + d_{n+1}) = 0 \quad (n \geq 1)
\end{array}\right.\label{sysd} \, ,
\eeq
where we have used a few handy multi-angle trigonometric relations such as
\beq
&2\cos 2z \cos 2nz = \cos 2(n+1)z + \cos 2(n-1)z \; (n\geq 1) \, ,\nonumber\\
&2\cos 2z \sin 2nz = \sin 2(n+1)z + \sin 2(n-1)z \; (n\geq 1) \, ,\nonumber\\
&2\cos 2z \cos (2n+1)z = \cos (2n+3)z + \cos (2n-1)z \; (n\geq 0) \, ,\nonumber\\
&2\cos 2z \sin (2n+1)z = \sin (2n+3)z + \sin (2n-1)z \; (n\geq 0) \, .\nonumber
\eeq

\begin{figure}
\centering
\includegraphics[width=0.7\textwidth,height=0.35\textheight]{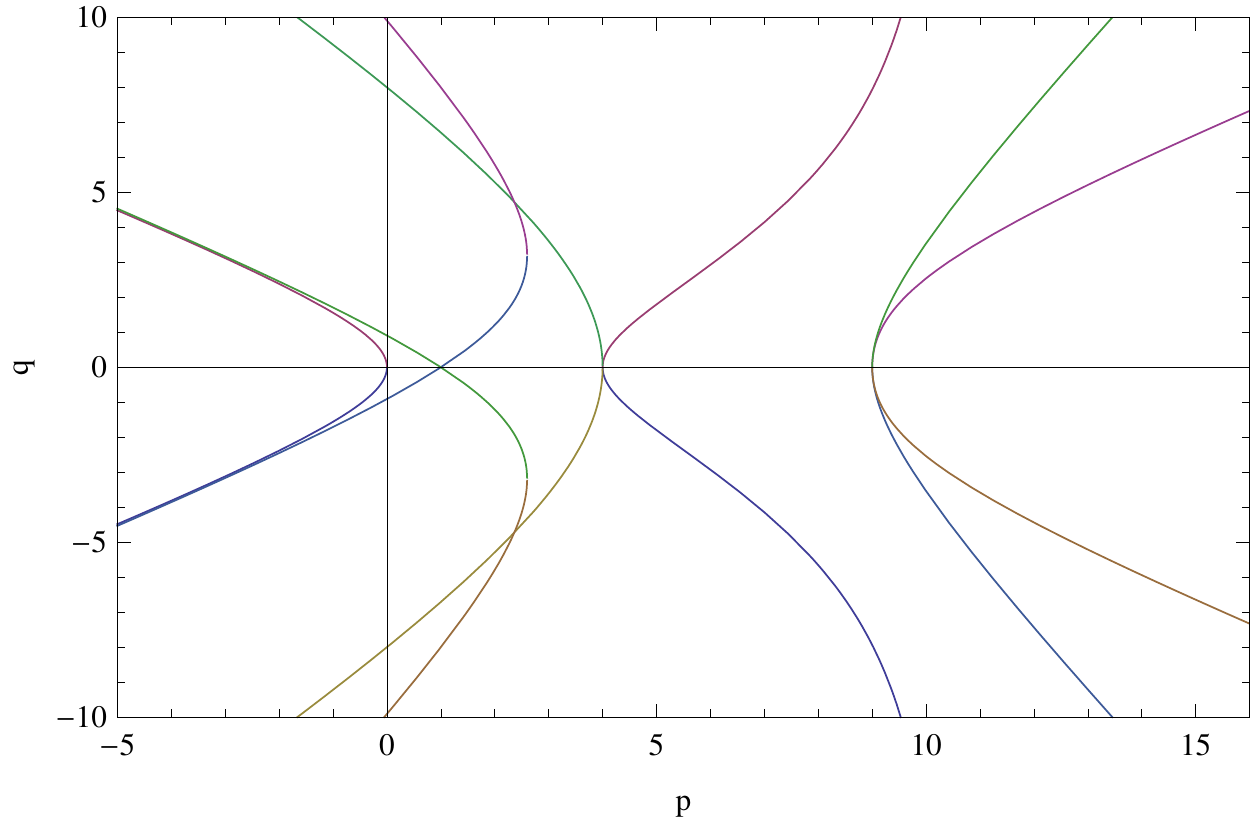}
\caption{Stability chart: the bands alternate between stable and resonant regions, from left to right or vice-versa, beginning with a stable one on either side.}
\label{chartfig}
\end{figure}

The systems~(\ref{sysa}), (\ref{sysb}), (\ref{sysc}), and (\ref{sysd}), must have zero determinants in order to possess non-trivial solutions for at least some of their coefficients $a_n$, $b_n$, $c_n$, and $d_n$. Therefore the equations
\beq
D_A =&
\begin{vmatrix}
  p & -q & 0 & \cdots \\
  -2q & p-4 & -q & \ddots \\
  0 & -q & p-16 & \ddots \\
  \vdots & \ddots & \ddots & \ddots
\end{vmatrix}
&= 0 \label{deta} \, ,\\
D_B =&
\begin{vmatrix}
  p-4 & -q & 0 & \cdots \\
  -q & p-16 & -q & \ddots \\
  0 & -q & p-36 & \ddots \\
  \vdots & \ddots & \ddots & \ddots
\end{vmatrix}
&= 0 \label{detb} \, ,\\
D_C =&
\begin{vmatrix}
  1-p+q & q & 0 & \cdots \\
  q & 9-p & q & \ddots \\
  0 & q & 25-p & \ddots \\
  \vdots & \ddots & \ddots & \ddots
\end{vmatrix}
&= 0 \label{detc} \, ,\\
D_D =&
\begin{vmatrix}
  1-p-q & q & 0 & \cdots \\
  q & 9-p & q & \ddots \\
  0 & q & 25-p & \ddots \\
  \vdots & \ddots & \ddots & \ddots
\end{vmatrix}
&= 0 \label{detd} 
\, ,
\eeq
must be satisfied (not simultaneously).  Each time one of those is satisfied we have identified a solution with the required period, that is, a boundary between stable and unstable regions in the $(p,q)$ plane.  We will confine ourselves to the case where $(p,q)\in \mathbb{R}$.  The stability chart is depicted in figure~\ref{chartfig}.

\paragraph{Characteristic exponents:}
We now seek some more solutions in the form (Floquet's)
\beq\label{floqf}
F_\nu(z) = e^{i\nu z} P(z) \, , \qquad F_\nu(-z) = e^{-i\nu z} P(-z) \, ,
\eeq
where $P$ is a regular oscillating function of period $\pi$.  Thanks to Floquet's theorem, as long as $\nu \notin \mathbb{Z}$ the two solutions $F(z)$ and $F(-z)$ are linearly independent, and the general solution to~(\ref{mathieuagain}) can be written as
\beq\label{floqs}
y(z) = A F_\nu(z) + B F_\nu(-z) \, ,
\eeq
with $A$ and $B$ constants.  Again, as long as $\nu \notin \mathbb{Z}$ we can decompose Floquet's solutions as
\beq\label{floqe}
F_\nu(z) = \sum_{n=-\infty}^{+\infty} f_n e^{i(\nu + 2n)z} \, \text{and} \, F_\nu(-z) = \sum_{n=-\infty}^{+\infty} f_n e^{-i(\nu + 2n)z} \, .
\eeq
Obviously, the region of instability we are interested in is that for which $\nu=i\mu$ with $\mu$ real.  Using the same machinery as before we could in principle find the solutions for $\mu$ on the $(p,q)$ plane.  In practice, we plug the expansions~(\ref{floqe}) back in the differential equation, extract recursive equations among coefficients, and their corresponding determinants must vanish.  The resulting master recursive equation is the same for both solutions, and reads
\beq
\label{floqr1}
\sum_{n=-\infty}^{+\infty} \left\{ \left[(\mu - 2in)^2 + p\right] f_n - q(f_{n-1}+f_{n+1}) \right\} e^{\mp\mu z \pm 2inz} = 0 \, ,
\eeq
from which the recursive relation
\beq
\label{floqr2}
\left[(\mu - 2in)^2 + p\right] f_n - q(f_{n-1}+f_{n+1}) = 0 \, ,
\eeq
can be extracted.  In principle one then analyses the determinant as before, reading
\beq
D_F^{(N)}(\mu) =
\begin{vmatrix}
  1 & \xi & 0 & \cdots \\
  \xi & 1 & \xi & \ddots \\
  0 & \xi & 1 & \ddots \\
  \vdots & \ddots & \ddots & \ddots
\end{vmatrix}_{N \times N}
= 0 \label{detf} \, ,
\eeq
where we define $\xi = q/\left[(i\mu + 2n)^2 - p\right]$, and $N\rar\infty$. This determinant can be recast as the summation
\beq\label{floqss}
D_F^{(N)} = \sum_{l=0}^{N} \alpha_l^N \xi^l \quad ; \quad \alpha_l^N = \alpha_l^{N-1} - \alpha_{l-2}^{N-2} \; , \; \alpha_0^N = 1 \; , \; \alpha_1^N = 0 \, .
\eeq
As far as we are aware of, the recursive relation can not be solved: since we need $D_F \equiv \lim_{N\rar\infty} D_F^{(N)}$ then we are not able to obtain an exact expression for the determinant.  

\begin{figure}
\centering
\includegraphics[width=0.48\textwidth]{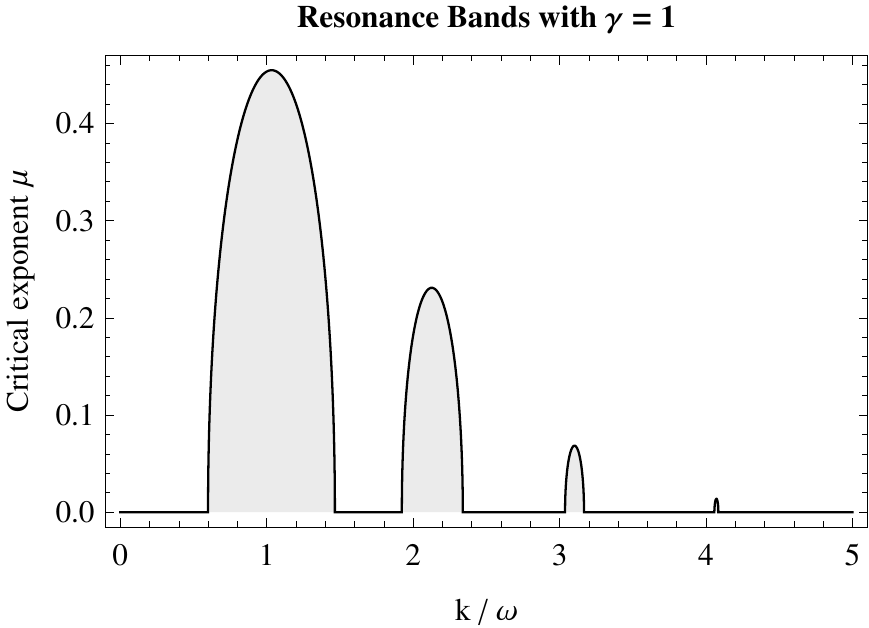}\hspace{3mm}
\includegraphics[width=0.48\textwidth]{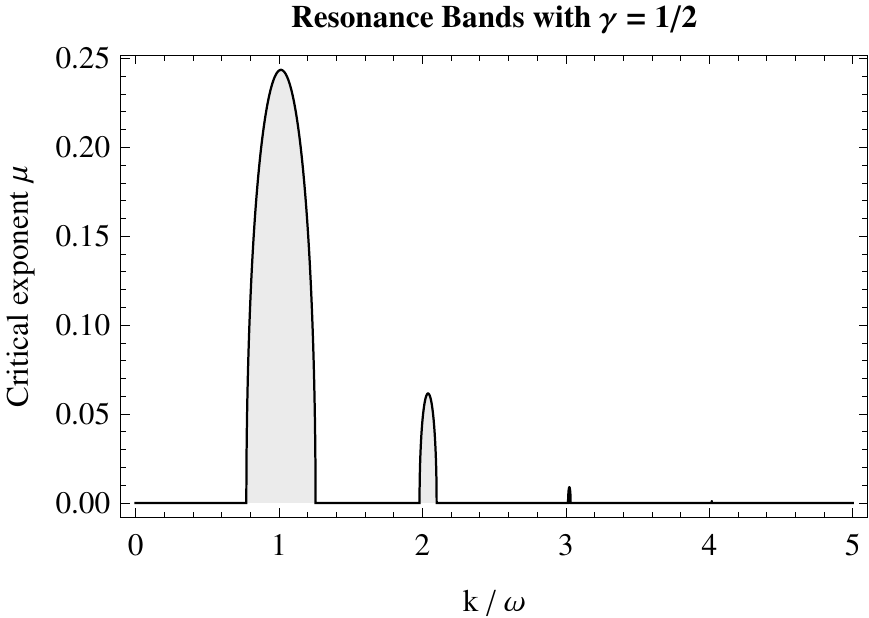}
\caption{Characteristic exponent and resonance bands.  The value of $q$ is $\sqrt p$ (left panel) or $\sqrt p / 2$ (right panel), that is, $\gamma =1$ (left) or $\gamma=1/2$ (right).}\label{bands}
\end{figure}

There is another path one can follow in order to have exact expressions which one can use to plot the characteristic exponent.  Let us take two solutions of Mathieu's equation which we call $y_1$ and $y_2$ such that
\beq
\left\{
\begin{array}{l}
  y_1(0) = 1 \\
  y_1'(0) = 0
\end{array}\right.\label{floqy1} \, ,\\
\left\{
\begin{array}{l}
  y_2(0) = 0 \\
  y_2'(0) = 1
\end{array}\right.\label{floqy2} \, .
\eeq
Now, the following relations can be proven:
\beq\label{floqm1}
\mu &=& -\frac{i}{\pi} \arccos\left[ y_1(\pi) \right] \, ,
\\ \label{floqm2}
\mu &=& -\frac{i}{\pi} \arccos\left[1 + 2\,y_1'(\pi/2)\, y_2(\pi/2) \right] \, .
\eeq
Since we ask that $\mu$ is real, we are only looking at purely imaginary values for the arccosine. In turn this means we are only interested in its argument being larger than +1 (or smaller than -1, in which case the real part is just $\pi$ and simply causes a sign flip in $F_\nu(z)$). The full explicit forms for $y_1$ and $y_2$ are combinations of sine and cosine Mathieu functions ($se$ and $ce$, respectively):
\beq\label{y1exp}
y_1(z) = \frac{ce'(a,q,0)se(a,q,z) - se'(a,q,0)ce(a,q,z)}{ce'(a,q,0)se(a,q,0) - se'(a,q,0)ce(a,q,0)} \, ,
\eeq
while
\beq\label{y2exp}
y_2(z) = \frac{ce(a,q,z)se(a,q,0) - se(a,q,z)ce(a,q,0)}{ce'(a,q,0)se(a,q,0) - se'(a,q,0)ce(a,q,0)} \, .
\eeq
After choosing a value for $q$, it is simple to plot the characteristic exponent as a function of $p$. In the context of the EM mode equation with an axial coupling to the inflaton these parameters correspond to $p=k^2/\omega^2$ and $q=\gamma k/\omega$. Figure~\ref{bands} shows examples of the resonance bands for this case.

%%%%%%%%%%%%%%%%%%%%%%%%%%%%%%%%%%%%%%%%%%%%%%%%%%%%%%%%%%%%%%%%%%%%%%%%%%%%%%%%%%%%%%
\bibliography{InflateResound}
\bibliographystyle{JHEP}
%%%%%%%%%%%%%%%%%%%%%%%%%%%%%%%%%%%%%%%%%%%%%%%%%%%%%%%%%%%%%%%%%%%%%%%%%%%%%%%%%%%%%%
\end{document}